\documentclass[10pt,letterpaper]{article}

\usepackage{opex3}
\usepackage{subfigure}
\usepackage{amsmath}
\usepackage{cite}
\usepackage[draft]{hyperref}

\begin{document}

\title{Terahertz time-domain spectroscopic ellipsometry: instrumentation and calibration}

\author{Mohammad Neshat$^*$ and N. P. Armitage}

\address{Department of Physics and Astronomy, Johns Hopkins University,\\ Baltimore, Maryland, 21218, USA}

\email{$^*$mneshat@jhu.edu} 



\begin{abstract}
We present a new instrumentation and calibration procedure for terahertz time-domain spectroscopic ellipsometry (THz-TDSE) that is a newly established characterization technique. The experimental setup is capable of providing arbitrary angle of incidence in the range of $15^\circ$--$85^\circ$ in the reflection geometry, and with no need for realignment. The setup is also configurable easily into transmission geometry. For this setup, we successfully used hollow core photonic band gap fiber with no pre-chirping in order to deliver a femtosecond laser into a THz photoconductive antenna detector, which is the first demonstration of this kind. The proposed calibration scheme can compensate for the non-ideality of the polarization response of the THz photoconductive antenna detector as well as that of wire grid polarizers used in the setup. In the calibration scheme, the ellipsometric parameters are obtained through a regression algorithm which we have adapted from the conventional regression calibration method developed for rotating element optical ellipsometers, and used here for the first time for THz-TDSE. As a proof-of-principle demonstration, results are presented for a high resistivity silicon substrate as well as an opaque Si substrate with a high phosphorus concentration. We also demonstrate the capacity to measure a few micron thick grown thermal oxide on top of Si. Each sample was characterized by THz-TDSE in reflection geometry with different angle of incidence.
\end{abstract}

\ocis{(040.2235) Far infrared or terahertz; (120.3940) Metrology; (120.5410) Polarimetry.} 


\section{Introduction}
Terahertz time-domain spectroscopy (THz-TDS) has tremendously grown with a wide range of applications  \cite{Grischkowsky_Oct1990, Fischer_2002, Mittleman_1998, Bilbro_2011}. It is now relatively routine to obtain complex (i.e. real and imaginary) spectral information, for instance the complex dielectric function, with absolute numerical values when performing measurements in a transmission geometry. However, transmission measurements are not possible on many materials and so far the technique has been difficult to apply to many metals, thick or highly doped semiconductors, coatings on thick substrates, substances in aqueous solution, and any otherwise opaque compound. Moreover, an outstanding technical problem with THz-TDS continues to be the determination of absolute spectral values when performing measurements in a reflection geometry  \cite{Nashima_Dec2001, Pashkin_2003}.  A reason why THz time-domain reflection measurements are challenging is that the time-domain technique rests on the ability to detect the relative amplitude and phase of a time-dependent electric field of a sample as compared to a reference. In transmission measurements, transmission through an aperture is used as a reference. For reflection based THz time-domain a simple mirror can not easily be used as a reference because its surface would need to be positioned within a fraction of a micron in exactly the same place as the sample so that the reflected phase is accurate. Such precise in situ positioning is challenging.

Ellipsometry is a well established technique in optical range whereby the measurement of the two orthogonal polarization components of light reflected at glancing incidence allows a complete characterization of a samples's optical properties at a particular frequency \cite{Fujiwara_2007}.  Typically, one measures the two orthogonal polarization components by a complete 360$^\circ$ characterization of the waves amplitude using rotating polarizers.  Importantly, ellipsometry obviates the need for measurement against a standard reference sample, and so can provide reliable spectroscopic information even when surface morphology is unknown, of marginal quality and/or a reference is unavailable. It is also self-referencing, so signal to noise ratios can be very good, as source fluctuations are divided out.  In order to overcome the technical problems mentioned above for THz reflectivity, ellipsometry techniques have been recently revisited for the terahertz range by using a backward wave oscillator source and a Golay cell power detector \cite{Hofmann_Feb2010}. There have been a number of attempts to extend ellipsometry to far-infrared frequencies using conventional Fourier Transform Spectroscopy technology \cite{Roseler_1990, Barth_1993, Bremer_Feb1992, Bernhard_2004, Kircher_Apr1997}. Unfortunately the lack of sufficiently intense sources (in addition to the calibration issues we confront here) has meant that such efforts have been challenging, although synchrotron based efforts have made some important contributions in this regard \cite{Bernhard_2004, Kircher_Apr1997}.   Generally these studies are limited to even higher frequencies ($>4$ THz) than we are here.

Combining ellipsometry with THz-TDS leads to a new technique called terahertz time-domain spectroscopic ellipsometry (THz-TDSE), in which a (sub)picosecond pulse with known polarization state is used as a probe to illuminate the sample, and then the modified polarization state by the sample is detected upon reflection or transmission. Unlike conventional optical ellipsometry, the reflected(transmitted) signal is detected coherently in the time-domain which allows one to obtain both amplitude and phase of the light in the two orthogonal directions. By transforming the time-domain data into the frequency domain through Fourier analysis, it is possible to extract ellipsometric parameter spectra similar to the standard optical spectroscopic ellipsometry. However, it should be noted that the instrumentation, signal analysis and calibration methods in THz-TDSE would differ from those in the standard optical ellipsometry, and all need to be revised accordingly. M. Hangyo and his coworkers, in their pioneering works, have demonstrated the potential of THz-TDSE for measuring the complex optical constants of a Si wafer with low resistivity, the soft-mode dispersion of SrTiO$_3$ bulk single crystals and the dielectric constants of doped GaAs thin films \cite{Nagashima_2001, Matsumoto_2009, Matsumoto_2011}. There are very few other reports on using the THz-TDSE \cite{Rubano_2012} technique, however, there has been recent work in developing high resolution THz polarimetry \cite{Shimano_2002, Castro-Camus_May2007, Dong_2009, Mechelen_2011, Hancock_2011, Aguilar_2012, Morris_May2012, George_2012, Yasumatsu_Jul2012} for material characterization. None of the proposed THz-TDSE experimental setups can provide an easy way of changing the angle of incidence without tedious work of optical/terahertz realignments. Moreover, less attention has been paid on the compensation of the non-idealities of the optical components and the alignments in THz-TDSE through a calibration scheme. Such compensation and calibration is essential in conventional optical range ellipsometry for accurate measurements, and as will show below is no-less essential in THz-TDSE.

In this paper, we present a new experimental setup for THz-TDSE in which the incidence angle in the reflection mode can change very easily in the range of $15^\circ-85^\circ$ with no need for any realignment. The same setup can be transformed into transmission mode with the same ease. Moreover, we propose a calibration scheme that can compensate for the non-ideality of the polarization response of the THz photoconductive antenna detector as well as that of wire grid polarizers used in the setup. In our calibration scheme, the ellipsometric parameters are obtained through a regression algorithm which we have adapted from the conventional regression calibration method developed for rotating element optical ellipsometers, and used here for the first time for THz-TDSE. As a demonstration, we present the characterization results for a high resistivity silicon substrate and a highly phosphorus doped Si substrate.  We also demonstrate the capacity to measure a few micron thick grown thermal oxide on top of Si. Each sample was characterized with different angle of incidence.


\section{Instrumentation}
\begin{figure}[htbp]
\centering
\subfigure[]{
   \includegraphics[width=11cm] {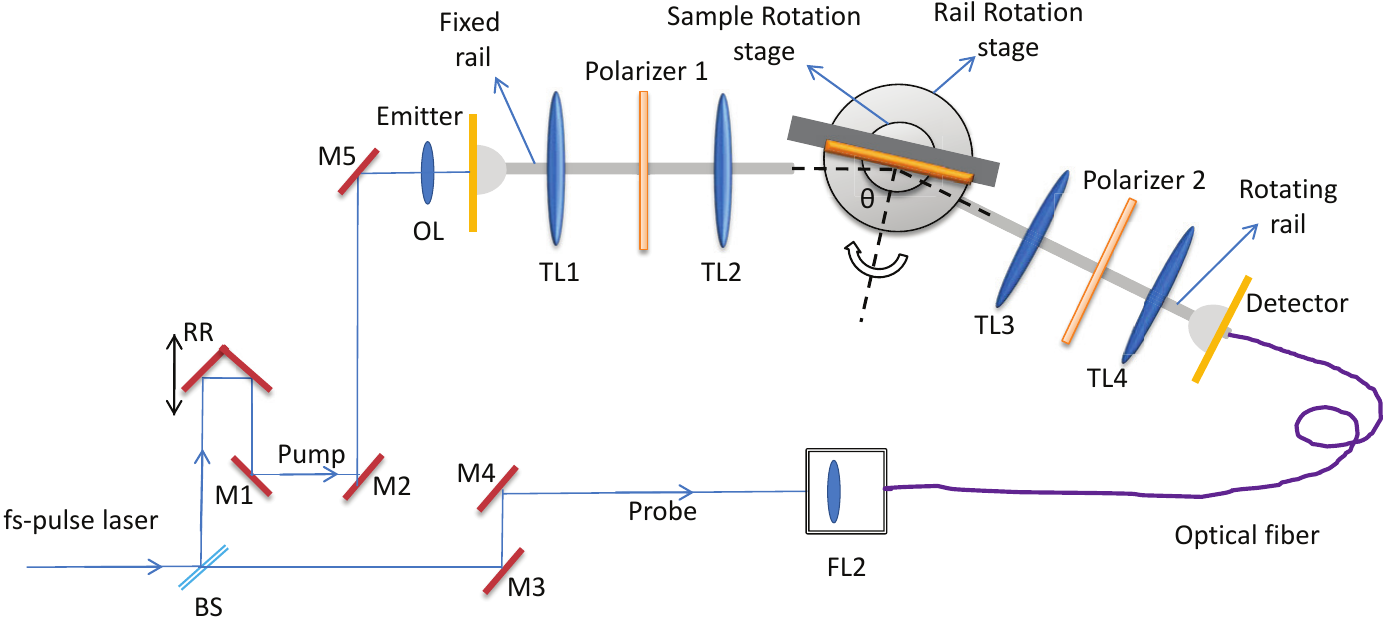}
}\\
\subfigure[]{
   \includegraphics[width=3.5cm] {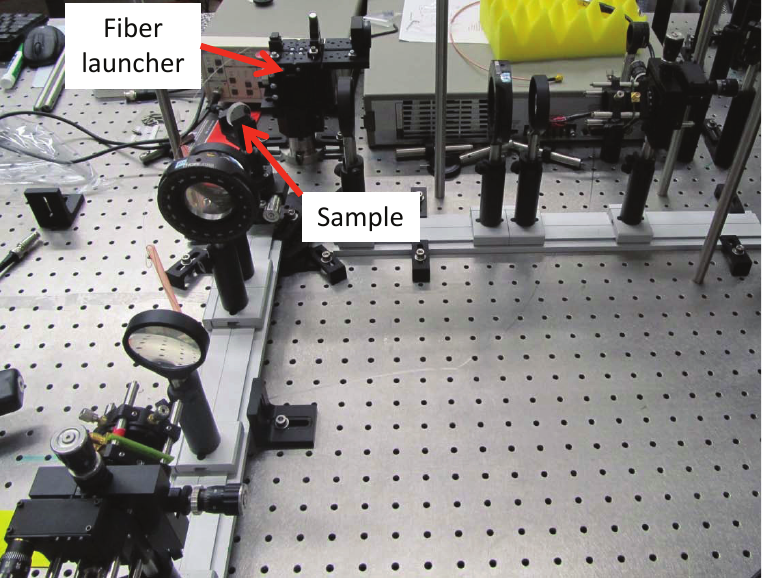}
}
\subfigure[]{
   \includegraphics[width=7.5cm] {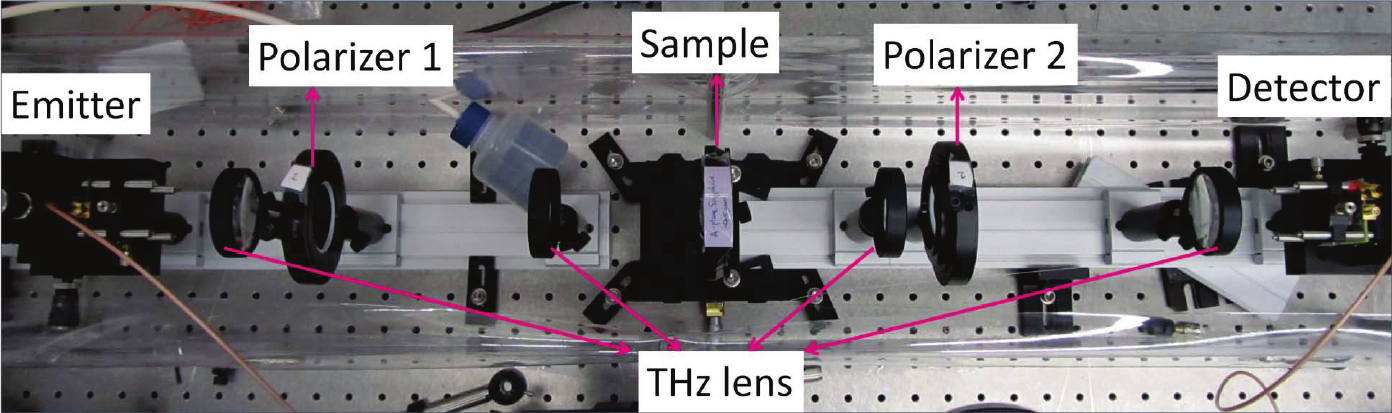}
}
\caption{(a) Schematic of the THz-TDSE setup with variable incidence angle in reflection mode that is also configurable in transmission mode (M, mirror; RR, retro-reflector; BS, optical beam splitter; OL, optical lens; TL, terahertz
lens), (b) Lab setup configured in reflection mode at $45^\circ$ incidence angle and (c) transmission mode.}\label{Setup}
\end{figure}

Figure \ref{Setup} illustrates our terahertz time-domain spectroscopic ellipsometry setup. In this system, the terahertz components are arranged on two straight arms using optical rails; one arm with THz emitter is fixed on the optical table, whereas the other arm with THz detector can rotate around a center point where the sample is placed. The sample sits on a second rotation stage. Such a configuration provides variable incidence angle \mbox{($15^{\circ}<\theta<85^\circ$)} in reflection mode, and can be easily configurable in transmission mode by aligning the arms along a straight line (Fig. \ref{Setup}(c)).

The setup uses an $8f$ confocal geometry with terahertz lenses made of \mbox{poly-4-methyl-pentacene-1} (TPX), which is a terahertz- and optically transparent material. Terahertz lenses have 50.8 mm clear aperture diameter and 100 mm focal length, and are less prone to misalignments
and polarization distortion as compared to off-axis parabolic
mirrors. The terahertz beam profile can be approximated as Gaussian. The sample is placed at the focal point of the terahertz lenses where an effective flat phase front exist at the beam waist. Based on the Gaussian beam assumption and 100 mm focal length for the THz lenses, the $f\#$ is approximately 9 for 0.2 THz. In the present case, the strong focusing and long wavelengths minimize the effects of a spread of incidence angles because an effective flat phase front exists at the sample position.  This is the opposite approach than has been used in previous infrared ellipsometers that were designed to work at high $f\#$ \cite{Kircher_Apr1997}. Two identical photoconductive dipole antennas with collimating substrate lens are used as THz emitter and detector. Two rotatable polarizers are placed in the collimated
beams immediately before the detector, and after the emitter as shown in Fig. \ref{Setup}. Polarizers
were wire grid with wire diameter and spacing of $10~\mu \textrm{m}$
and $25~\mu \textrm{m}$, respectively, and field extinction ratio of
$\sim 40:1$ at 1 THz.

The laser source is an 800 nm Ti:sapphire
femtosecond laser with pulse duration of \mbox{$<20$ fs} and
85 MHz repetition rate, which is divided into pump and
probe beams. The pump beam is guided and focused onto the gap of the emitter photoconductive antenna through free-space optics in the usual fashion. However, the probe beam is guided toward the detector antenna on the rotating arm by a 1 m long hollow core photonic band-gap fiber (HC-PBGF). Using a fiber-coupled THz detector facilitates the rotation of the detector arm with no need for realignment of the optics for each incidence angle. The advantage of HC-PBGF is that it  exhibits extremely low nonlinearity, high breakdown threshold, zero dispersion at the design wavelength, and negligible interface reflection \cite{Hensley_2008}. Near zero dispersion of such fibers elliminates the need for optical pulse pre-chirping, which makes the optical setup and the alignments considerably simpler than using convensional fibers with pre-chirping.

The fiber used in our setup is commercially available HC-800-2 from NKT Photonics. This fiber, fabricated by a stack-and-draw method, has a low-loss transmission band-gap of 100 nm centered around 820 nm. The core diameter is $7.5~\mu$m, and the mode-field diameter is $5.5~\mu$m. Attenuation through the low-loss region is reported as $<250$ dB/km. The fiber is connected from one side to a fiber launcher with a focusing lens and a 3-axis positioner, and from the other side is connected to a fiber-coupled focusing optics that is attached to the THz detector compartment.

\begin{figure}[htbp]
\centering
\subfigure[]{
   \includegraphics[width=6.3cm] {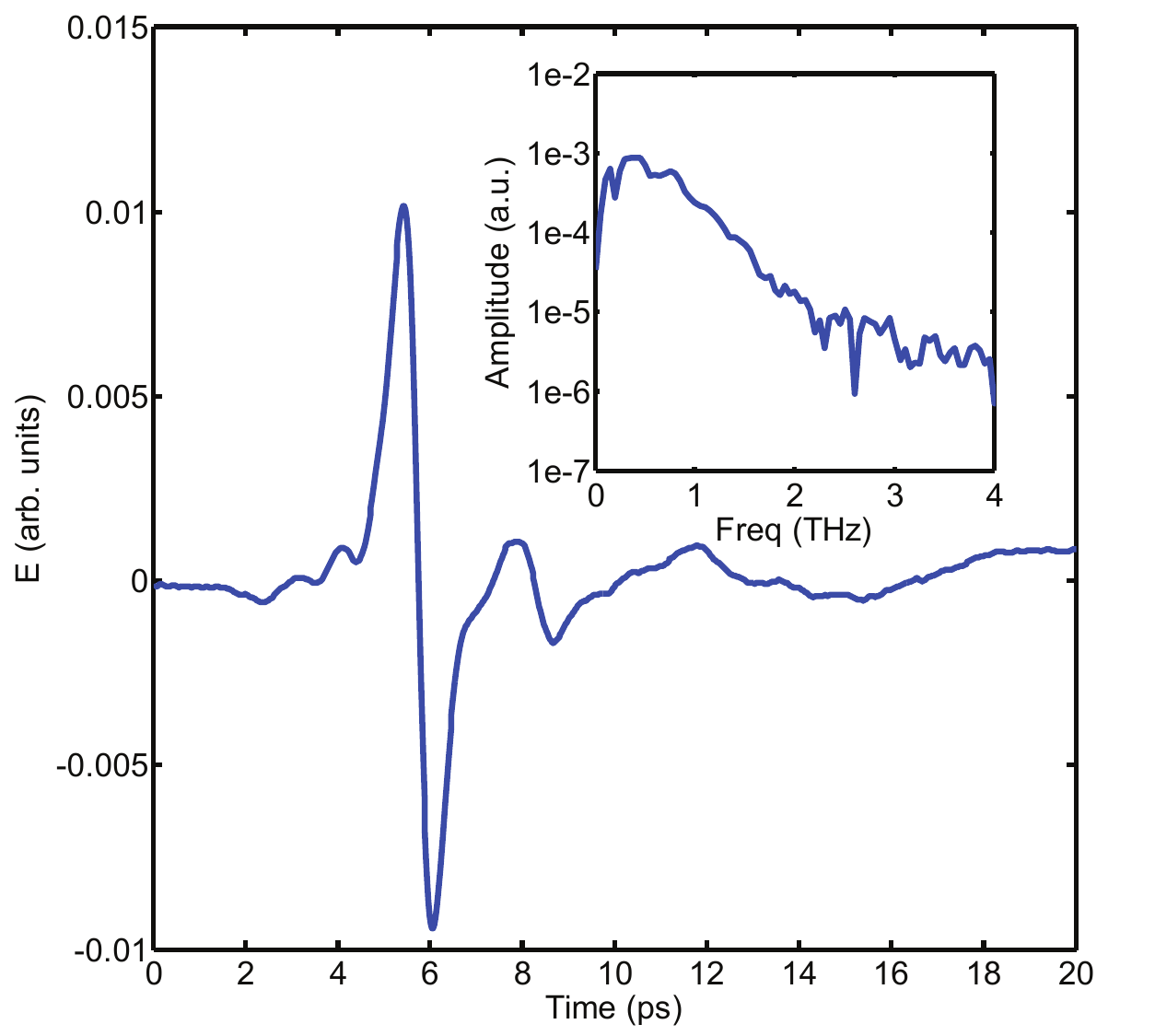}
}
\subfigure[]{
   \includegraphics[width=5.9cm] {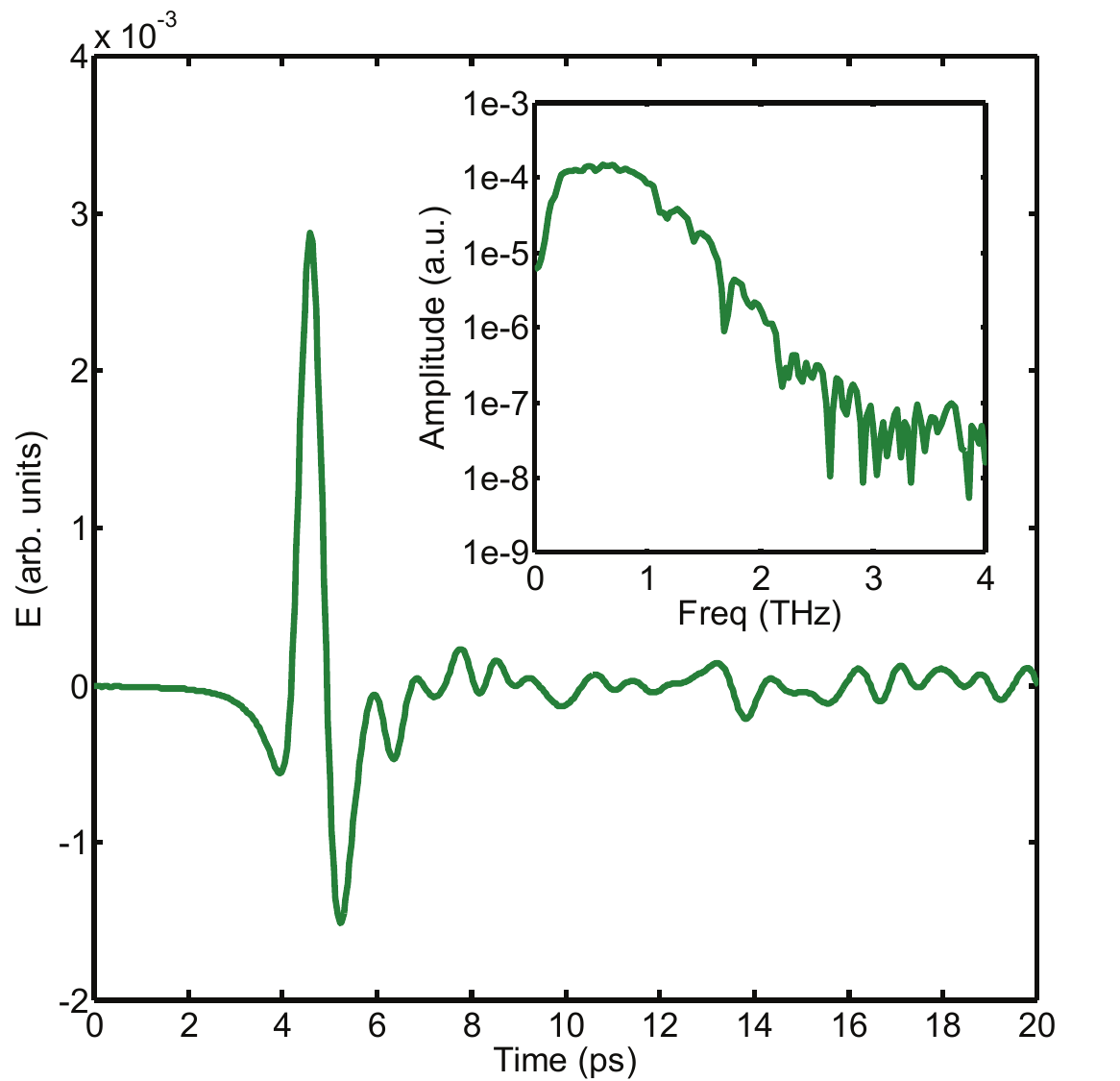}
}
\caption{Typical THz pulse detected in transmission mode when the probe beam is guided through (a) 1-m-long hollow core photonic band-gap fiber, (b) free-space optics. Insets show the corresponding pulse spectra.}\label{THzPuls}
\end{figure}

\begin{figure}[htbp]
\centering
\subfigure[]{
   \includegraphics[width=6cm] {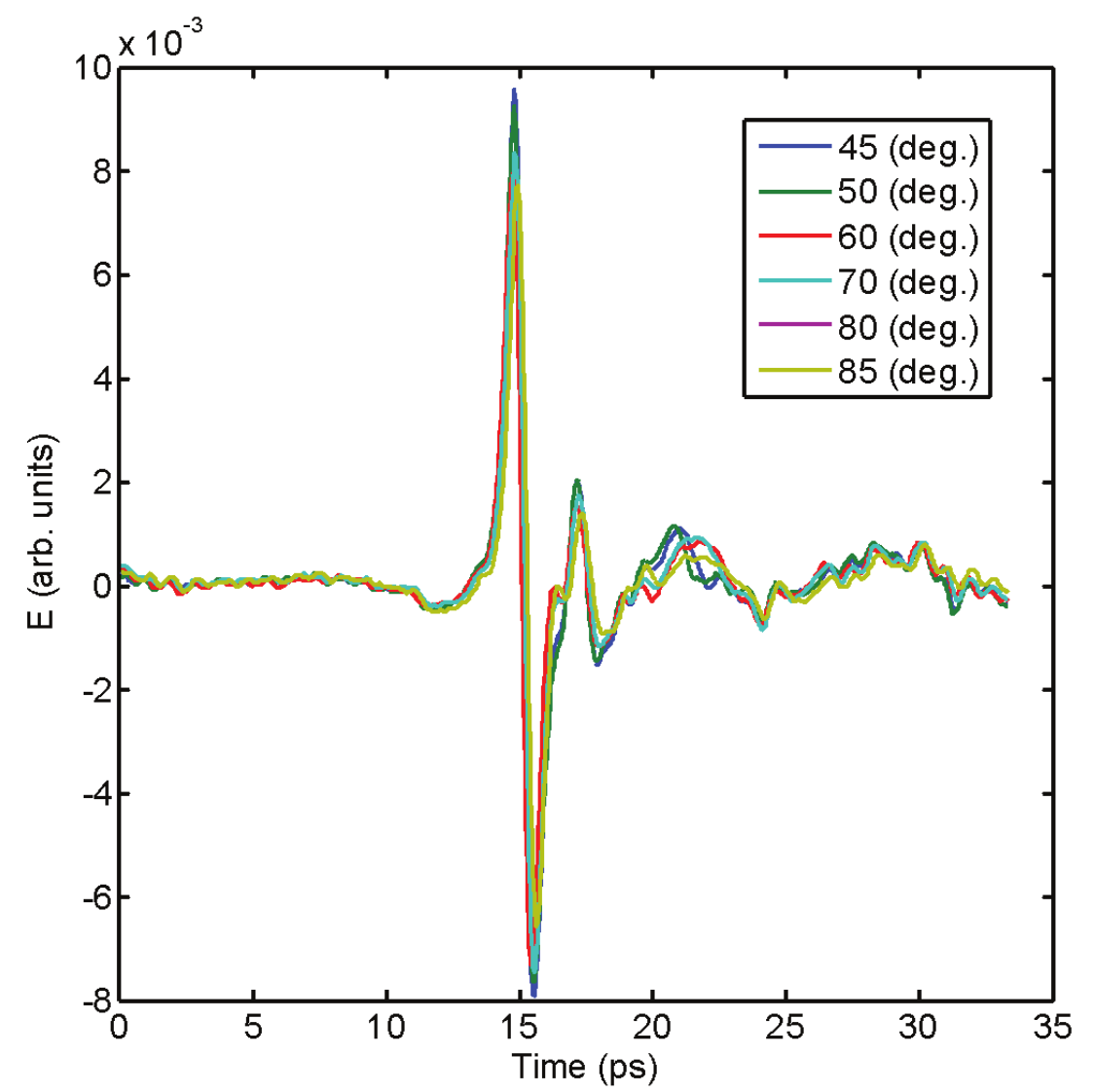}
}
\subfigure[]{
   \includegraphics[width=6cm] {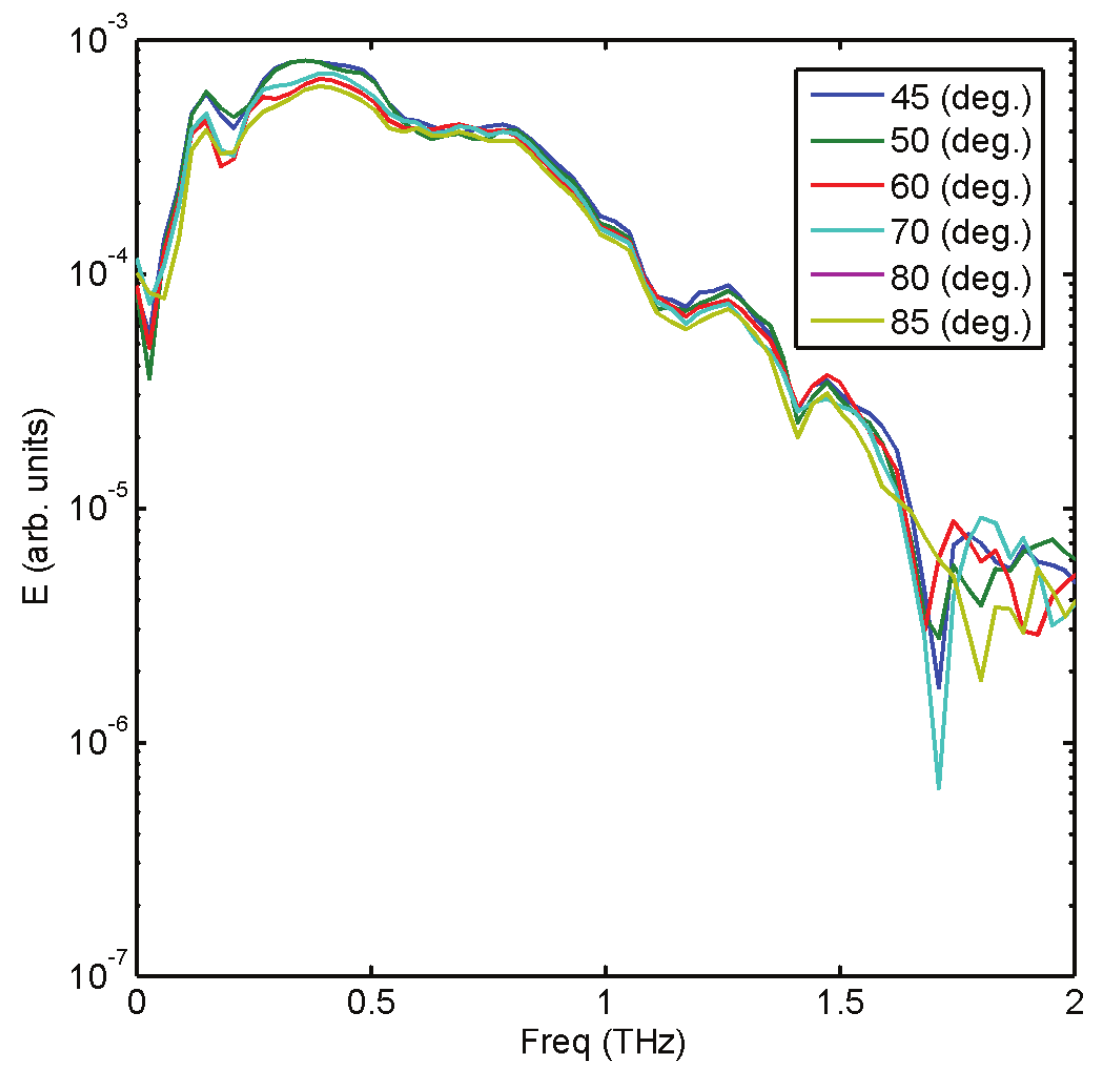}
}
\caption{(a) Time-domain THz pulse and (b) its corresponding spectrum in reflection mode for various incidence angles. A silver mirror was used as the reflector.}\label{PulsReflection}
\end{figure}

The temporal terahertz pulse is acquired by a lock-in amplifier during a time window within which the time delay between terahertz pulse and the sampling probe laser is swept by continuous movement of the retro-reflector. The pump beam is mechanically chopped, and the lock-in amplifier is synchronized to the chopping frequency. The acquired temporal signal is then taken into the frequency domain through a Fourier transform.  Figure \ref{THzPuls} shows a typical THz pulse detected in transmission mode without any sample. In Fig. \ref{THzPuls}, one can compare the time-domain pulse shape and its spectrum detected when the probe beam is delivered to the detector by a 1 m long hollow core photonic band-gap fiber vs. free-space coupling.  Figure \ref{PulsReflection} shows the THz pulse and its corresponding spectrum in reflection using fiber coupling for various incidence angles without any optics realignment after changing the incidence angle.   One can see that in this setup, any changes in the overall intensity at different angles are very small.  In the reflection measurements a silver mirror was used as a reflector.

\section{Formulation}
Real-valued ellipsometric parameter spectra $(\Psi,\Delta)$ are obtained as a result of the spectroscopic ellipsometry measurement. These parameters are related to the complex ratio of the polarization eigenstates corresponding to the \emph{p} and \emph{s} components of the electric field vector (\emph{p}- and \emph{s}-axis are defined perpendicular to the propagation direction, and parallel and perpendicular to the incidence plane, respectively). Specifically,  $\tan\Psi$ and $\Delta$ are defined as the absolute value and the phase of the complex ratio, respectively, when the incident polarization state is linear and at $45^\circ$ azimuthal angle \cite{Fujiwara_2007}. Once $\Psi$ and $\Delta$ spectra are obtained from experiments, the optical constants and layer thicknesses can be generally extracted by fitting proper dielectric functions and optical models to the measured $\Psi$ and $\Delta$ parameters. In a specific case of a bulk sample the complex refractive index ($n-jk$) can be obtained directly from the ellipsometric parameters using the expressions \cite{Born_1999}

\begin{align}\label{bulk_Eq}
n^2-k^2&=\sin^2\theta\bigg[1+\frac{\tan^2\theta(\cos^22\Phi-\sin^22\Phi\sin^2\Delta)}{(1+\sin 2\Phi\cos\Delta)^2}\bigg],\\\nonumber
2nk&=\sin^2\theta\frac{\tan^2\theta\sin 4\Phi\sin\Delta}{(1+\sin 2\Phi\cos\Delta)^2},
\end{align}

\noindent where $\theta$ is the incidence angle, and $\tan\Phi=1/\tan\Psi$.

\begin{figure}[htbp]
\centering
\includegraphics[width=11cm] {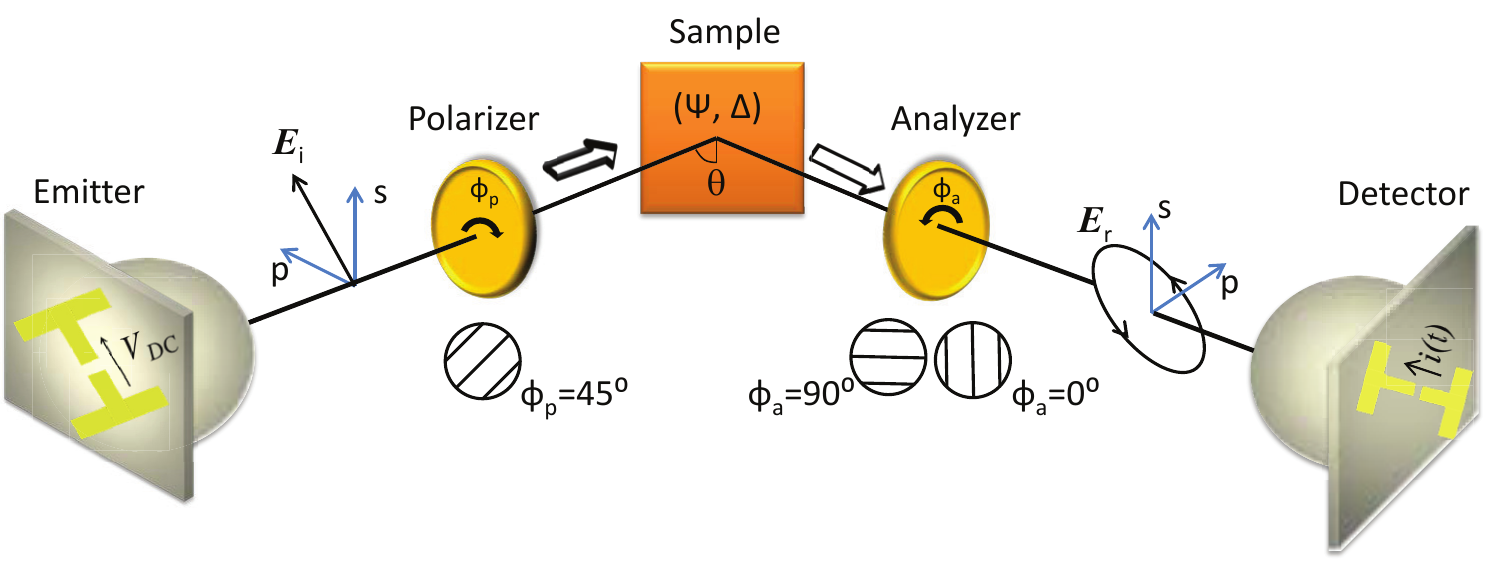}
\caption{Schematic of the optical configuration of THz-TDSE in reflection mode.}\label{Ellipsometry}
\end{figure}

Figure \ref{Ellipsometry} illustrates a schematic of the optical configuration of the THz-TDSE in reflection mode. A biased photoconductive dipole antenna (Emitter) generates the incident field \mbox{$\mathbf{E}_i=[E_{ip}~E_{is}]$} with close to linear polarization state and azimuthal angle around $45^\circ$. After passing through the polarizer, the polarization state becomes highly linear at the azimuthal angle of the polarizer $\phi_p$. The reflected polarization state is then modified according to the ellipsometric parameters $(\Psi,\Delta)$ of the sample. The \emph{p}- and \emph{s}-component of the reflected polarization state are detected by setting the azimuthal angle of the analyzer at $\phi_a=0^\circ$ and $90^\circ$, respectively.

In frequency domain, the Jones vector of the light beam at the THz detector, $\mathbf{E}_r=[E_{rp}~E_{rs}]$ as shown in Fig. \ref{Ellipsometry}, is determined by the product
of the Jones matrices of the polarizer $\mathbf{J}_P$, sample $\mathbf{J}_S$, and analyzer $\mathbf{J}_A$ as \cite{Fujiwara_2007}

\begin{equation}\label{matrix}
\mathbf{E}_r=\mathbf{R}(-\phi_a)\mathbf{J}_A\mathbf{R}(\phi_a)\mathbf{J}_S\mathbf{R}(-\phi_p)\mathbf{J}_P\mathbf{R}(\phi_p)\mathbf{E}_i,
\end{equation}

\noindent where $\mathbf{R}(\phi)$ is the rotation matrix with angle $\phi$. The Jones matrices for the sample and the polarizer/analyzer are given as \cite{Fujiwara_2007}

\begin{equation}
\mathbf{J}_s=\begin{bmatrix}
\textrm{sin}(\Psi)\textrm{exp}(j\Delta) & 0\\
0 & \textrm{cos}(\Psi)
\end{bmatrix},~~~
\mathbf{J}_{P(A)}=\begin{bmatrix}
1 & 0\\
0 & \eta_{P(A)}\textrm{exp}(j\delta_{P(A)})
\end{bmatrix},\label{Jones}
\end{equation}

\noindent where $\eta_{P(A)}$ is the extinction ratio of the polarizer(analyzer), defined as the ratio of the transmitted electric field when the polarizer transmission axis is perpendicular and parallel to the electric field direction, and $\delta_{P(A)}$ is the phase retardance. For an ideal polarizer/analyzer the extinction ratio is zero ($\eta_{P(A)}=0$).

In practice, the response of the THz detector is far from ideally linear polarization due to the non-ideality of the antenna geometry, and optical and terahertz alignments \cite{Gong_Sep2011,Castro-Camus_May2007}. Therefore, the polarization response of the THz detector should be taken into account by introducing a polarization state vector $\mathbf{P}_d=[P_p~P_s]$, which is generally frequency dependent. In the frequency domain, the spectrum of the output signal of the photoconductive antenna detector can be expressed by an inner product as \cite{Neshat_Jun2012}

\begin{equation}
I=\mathbf{P}_d(\omega)\cdotp\mathbf{E}_r,\label{Eq1}
\end{equation}

\noindent where $I$ is the Fourier transform (spectrum) of the THz detector output signal, $\mathbf{E}_r$ is the electric field vector impinging on the detector as shown in Fig. \ref{Ellipsometry}, and $\omega$ is the angular frequency. We define a complex parameter $\rho$ as the ratio of the detector output spectrum when the transmission axis of the analyzer is along \emph{p}- and \emph{s}-axis which corresponds to $\phi_a=0^\circ$ and $90^\circ$, respectively, as

\begin{equation}
\rho_{\textrm{mod}(\text{exp})}(\Psi,\Delta,\phi_p,\eta_{P(A)},\delta_{P(A)})=\frac{I_{\textrm{mod}(\textrm{exp})}(\phi_a=0^\circ)}{I_{\textrm{mod}(\textrm{exp})}(\phi_a=90^\circ)},\label{Eq2}
\end{equation}

\noindent where $I_{\textrm{mod}}$ and $I_{\textrm{exp}}$ refer to the detector output spectrum calculated through the model described by Eqs. (\ref{matrix})-(\ref{Eq1}), and obtained directly from the experiment, respectively. From Eqs. (\ref{matrix})-(\ref{Eq1}), it is clear that the $\rho$-parameter is a function of the sample ellipsometric parameters $(\Psi,\Delta)$, azimuthal angle of the polarizer $\phi_p$, and the extinction ratio and phase retardance of both polarizer and analyzer. It is worth noting that under ideal conditions, i.e. $\eta_{P(A)}=0$ and \mbox{$\mathbf{P}_d=[1/\sqrt{2}~~~1/\sqrt{2}]$}, and for $\phi_p=45^\circ$, the ellipsometric parameters of the sample can be directly extracted from the $\rho$-parameter as

\begin{equation}
\Psi=\textrm{tan}^{-1}|\rho|,~~~\Delta=arg(\rho),\label{Eq3}
\end{equation}

\noindent where $arg(.)$ denotes the phase operator.
\section{Measurement and Calibration Procedure}
In order to accurately measure the ellipsometric parameters ($\Psi,\Delta$) and consequently the optical constants through THz-TDSE, one needs to know the polarization state vector of the detector $\mathbf{P}_d$, extinction ratio and phase retardance of the polarizer/analyzer as well as their azimuthal offset angles $\Delta\phi_p/\Delta\phi_a$ with respect to the \emph{p}-axis. Azimuthal offset angles may arise due to the uncertainty in the polarizer (analyzer), or the misalignment of the components (e.g. surface tilt of the sample), and can be easily modeled by replacing $\mathbf{R}(\phi_{p(a)})$ with $\mathbf{R}(\phi_{p(a)}+\Delta\phi_{p(a)})$ in Eq. (\ref{matrix}).

The complex vector $\mathbf{P}_d$ can be obtained experimentally over the desired frequency range through a calibration scheme reported elsewhere \cite{Neshat_Jun2012}. Using a gold mirror as the sample in Fig. \ref{Ellipsometry}, and by adapting the calibration scheme in \cite{Neshat_Jun2012} for the reflection mode, it is easy to show that

\begin{equation}
\mathbf{P}_d=\begin{bmatrix}
P_p\\[0.4em]
P_s
\end{bmatrix}=\begin{bmatrix}
~\sqrt{\frac{1+s}{2}}+\sqrt{\frac{1-s}{2}}\exp{(j\delta_d)}\\[0.4em]
-\sqrt{\frac{1+s}{2}}+\sqrt{\frac{1-s}{2}}\exp{(j\delta_d)}
\end{bmatrix},\label{Pd}
\end{equation}

\noindent where
\begin{align}
s&=\frac{1-|\kappa|^2}{1+|\kappa|^2},\\
\delta_d&=arg(\kappa),\\
\kappa&=\frac{I_{\textrm{exp}}(\phi_a=0^\circ)-I_{\textrm{exp}}(\phi_a=90^\circ)}{I_{\textrm{exp}}(\phi_a=0^\circ)+I_{\textrm{exp}}(\phi_a=90^\circ)}.
\end{align}

Once $\mathbf{P}_d$ is known from Eq. (\ref{Pd}) and through the reflection measurement from a gold mirror, the ellipsometric parameters are obtained through a regression algorithm which we have adapted from the conventional regression calibration method developed by B. Johs for rotating element optical ellipsometers \cite{Johs_1993}, and used here for the first time for THz-TDSE. In our proposed regression algorithm, the $\rho$-parameter defined in Eq. \ref{Eq2} is obtained experimentally ($\rho_{\textrm{exp}}$) from an isotropic sample at one or several different azimuthal angles of the polarizer ($\phi_{pi}, i=1,..., N$). Then, the $\rho$-parameter from the model ($\rho_{\textrm{mod}}$) described by Eqs. (\ref{matrix})-(\ref{Eq2}) is fitted to that from the experiment over the desired frequency range and simultaneously for all the polarizer azimuthal angles, using the Levenberg–-Marquardt nonlinear regression algorithm. We define an error function \emph{Err} based on least squares fitting as

\begin{align}\label{Err_eq}
Err=\sum_{\phi_{pi}}\sum_{\omega_j}&\big[|\rho_{\textrm{mod}}(\omega_j,\phi_{pi},\Psi,\Delta,...)|-|\rho_{\textrm{exp}}(\omega_j,\phi_{pi})|\big]^2+\\\nonumber
&\big[arg(\rho_{\textrm{mod}}(\omega_j,\phi_{pi},\Psi,\Delta,...))-arg(\rho_{\textrm{exp}}(\omega_j,\phi_{pi}))\big]^2.
\end{align}

It should be noted that all measurements and computations in this work are made in discrete time and frequency domains. In the regression algorithm, the ellipsometric parameters $\Psi$ and $\Delta$, as well as the polarizer/analyzer parameters $\eta_{P(A)}$, $\delta_{P(A)}$ and $\Delta\phi_{p(a)}$ are simultaneously found such that the error function defined in Eq. (\ref{Err_eq}) is minimized over the measurement frequency range ($\omega_j, j=1,..., M$) and for all measured data sets corresponding to different polarizer angles ($\phi_{pi}, i=1,..., N$). The advantage of this technique is that the polarizer/analyzer parameters do not need to be a priori knowledge, although reasonable initial values are necessary for the convergence of the regression algorithm. It was found that the best initial values for $\Psi$ and $\Delta$ are those obtained from Eq. (\ref{Eq3}) when $\rho=\rho_{\textrm{exp}}$.
\section{Experimental results and discussion}
As a demonstration, two different silicon samples were tested. The first sample was a \mbox{0.4-mm-thick} high resistivity Si substrate with polished surface. For this sample the reflected time-domain pulse was truncated right before the arrival of the second reflection in order to avoid Fabry-perot interferences due to the reflection from the back surface. The second sample was a 1-mm-thick highly phosphorus-doped Si substrate with nominal room temeprature DC resistivity of $0.0151~\Omega\textit{-}\textrm{cm}$.

\begin{figure}[htbp]
\centering
\subfigure[]{
   \includegraphics[width=5.5cm] {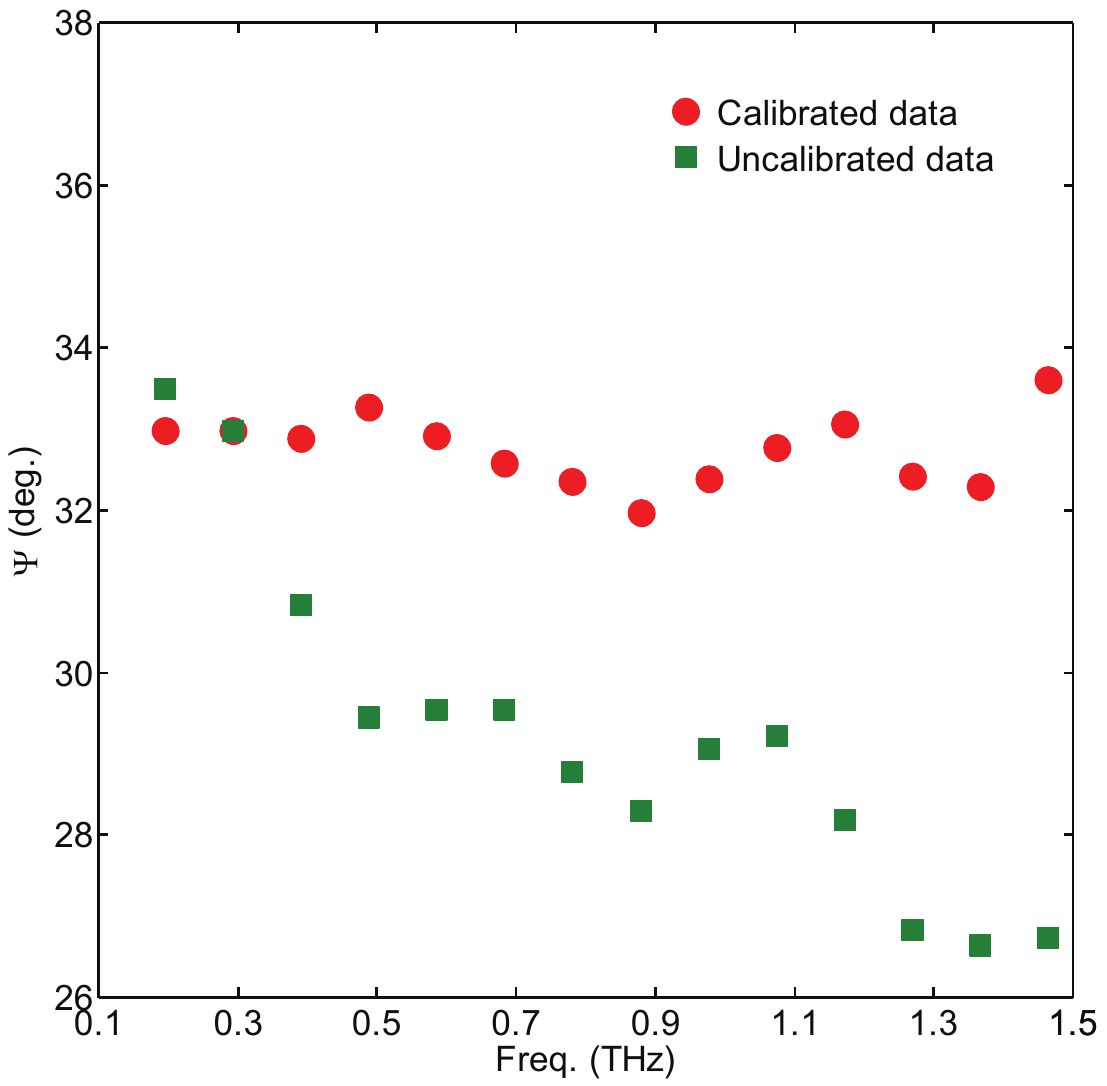}
}
\subfigure[]{
   \includegraphics[width=5.6cm] {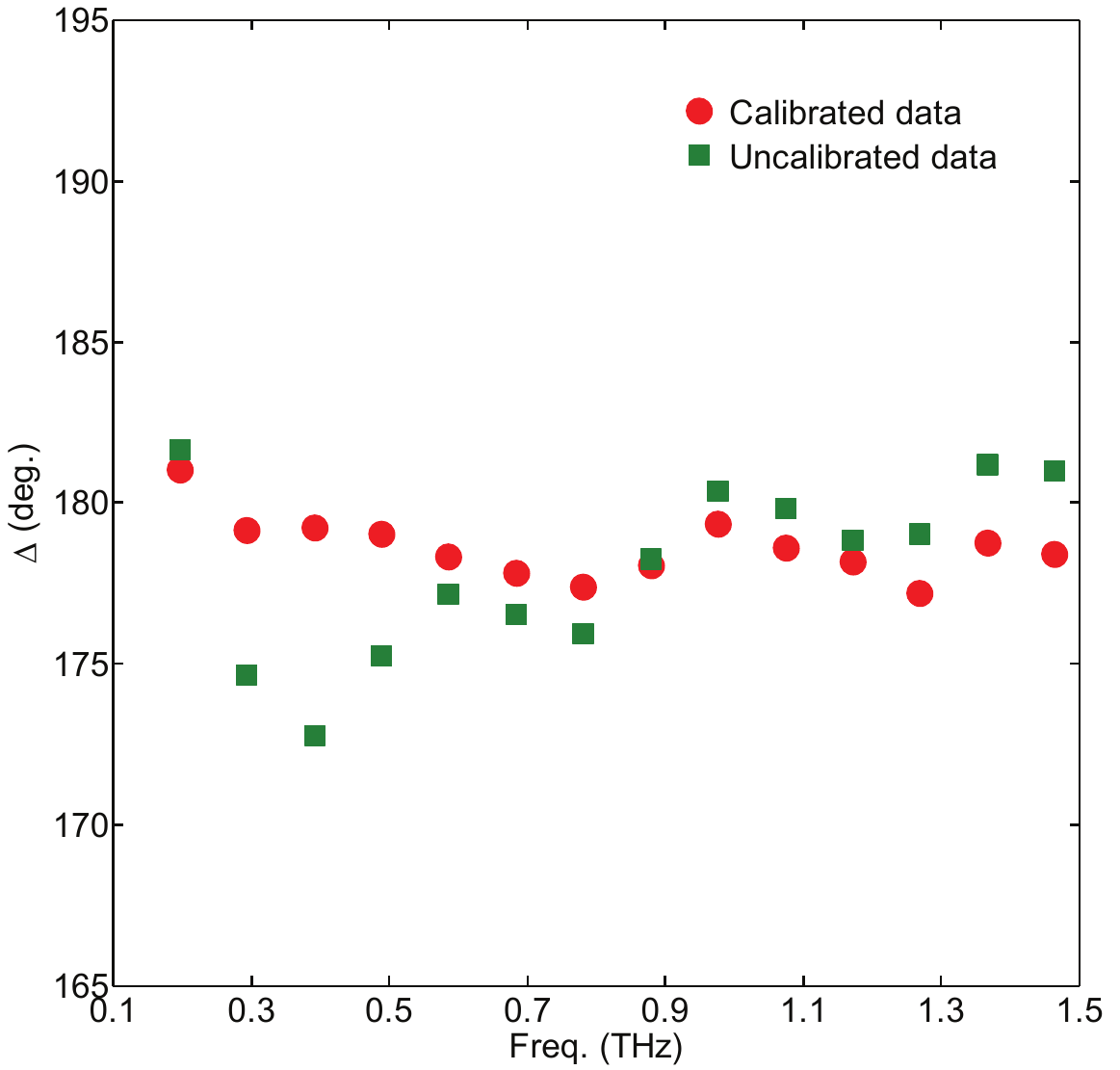}
}
\caption{Measured ellipsometric parameters (a) $\Psi$ and (b) $\Delta$ for the high resistivity silicon substrate. Square and circle marks present the ellipsometric parameters before and after calibrating out the effect of the THz detector and the polarizer/analyzer, respectively. The incidence angle was $45^\circ$.}\label{SiHR}
\end{figure}

Figure \ref{SiHR} shows the measured calibrated and uncalibrated ellipsometric parameters $\Psi$ and $\Delta$ for the high-resitivity sample. The calibrated plots are obtained by fitting the $\rho$-parameter to three sets of measured data corresponding to polarizer angles $\phi_{p1}=25^\circ$, $\phi_{p2}=45^\circ$ and $\phi_{p3}=65^\circ$. The uncalibrated plot is directly obtained from Eq. (\ref{Eq3}) using the data set corresponding to $\phi_{p2}=45^\circ$. An incidence angle of $45^\circ$ (away from the Brewster's angle) was used for this experiment so as to have appreciable signal in both the $s-$ and $p-$polarizations.

\begin{figure}[htbp]
\centering
\includegraphics[width=5.5cm] {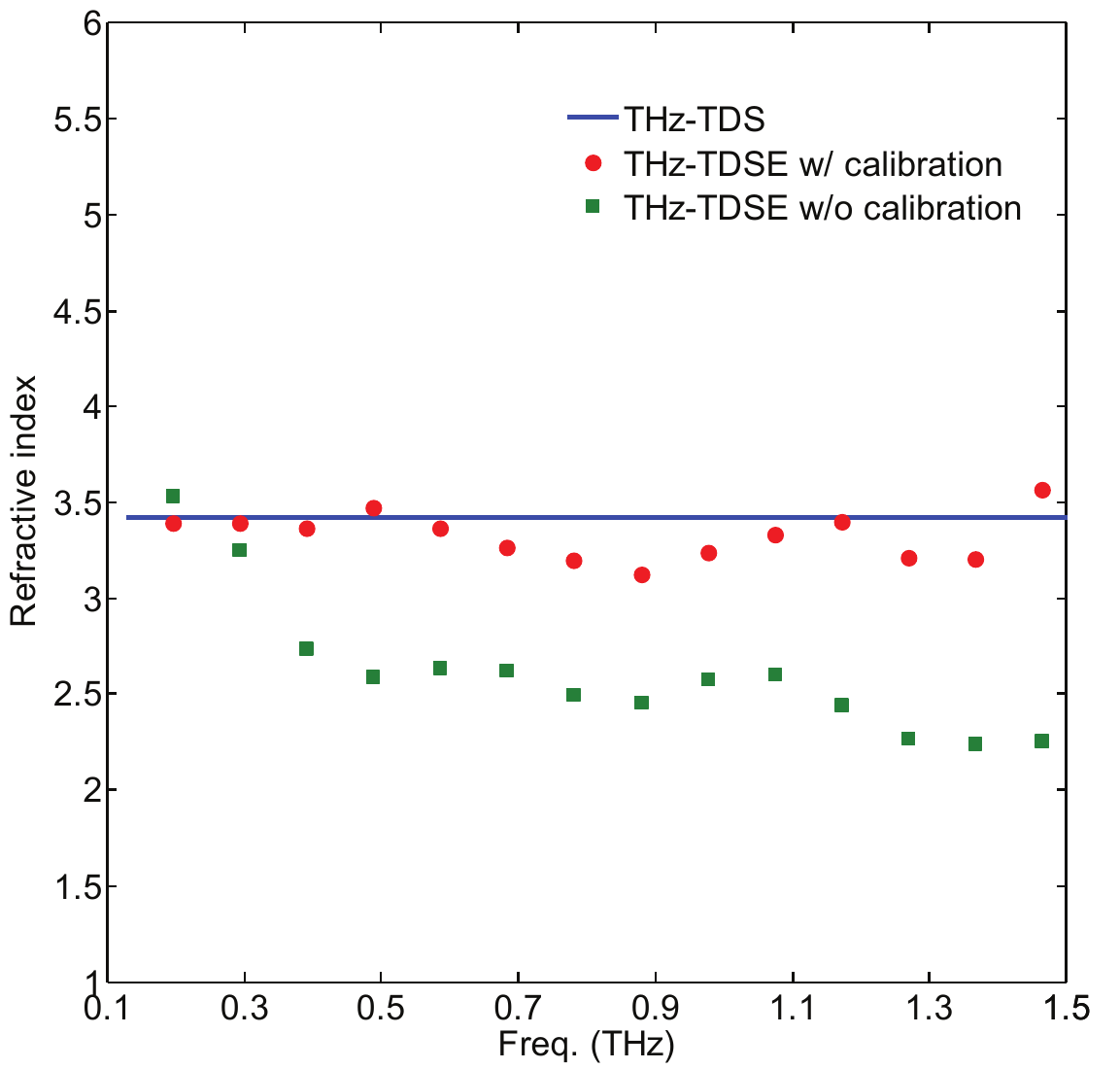}
\caption{Extracted refractive index of the high resistivity silicon substrate from the uncalibrated (square) and calibrated (circle) ellipsometric parameters. Solid line shows the data from the conventional transmission THz-TDS reported in \cite{Grischkowsky_Oct1990}.}\label{RI_SiHR}
\end{figure}

Figure \ref{RI_SiHR} compares the extracted refractive index, using Eq. (\ref{bulk_Eq}), from the uncalibrated and calibrated ellipsometric parameters shown in Fig. \ref{SiHR} with that from the conventional transmission THz-TDS reported in \cite{Grischkowsky_Oct1990}. It is evident that the calibration has substantially improved the measurement accuracy over the uncalibrated data.   The ellipsometric measurements do not have as much precision as the transmission experiments.   But this illustrates the important point that the power of ellipsometric measurements rests not on measuring dielectric properties of samples that could be measured in transmission, but in measuring samples that are otherwise opaque.

\begin{figure}[htbp]
\centering
\subfigure[]{
   \includegraphics[width=5.5cm] {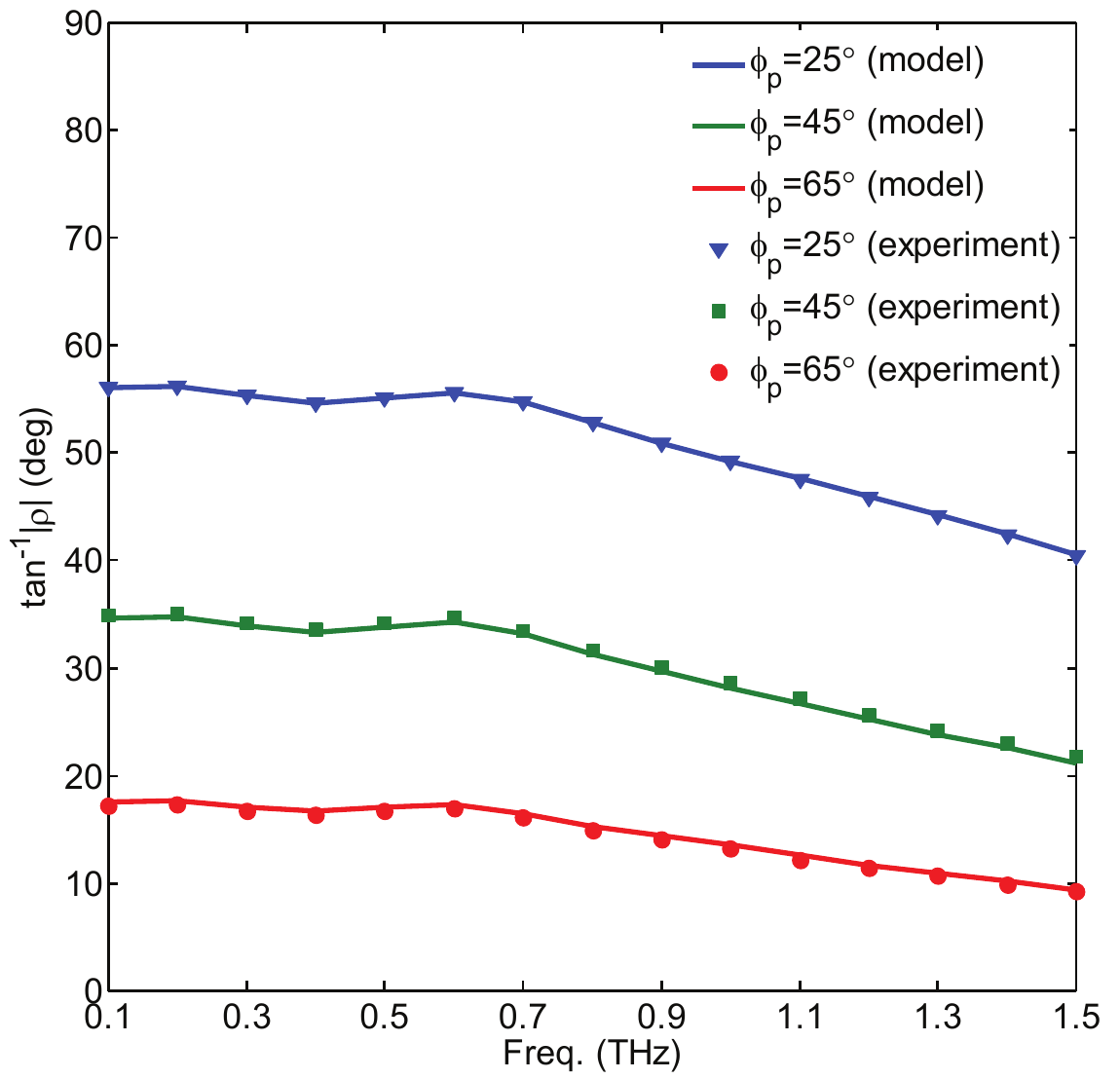}
}
\subfigure[]{
   \includegraphics[width=5.5cm] {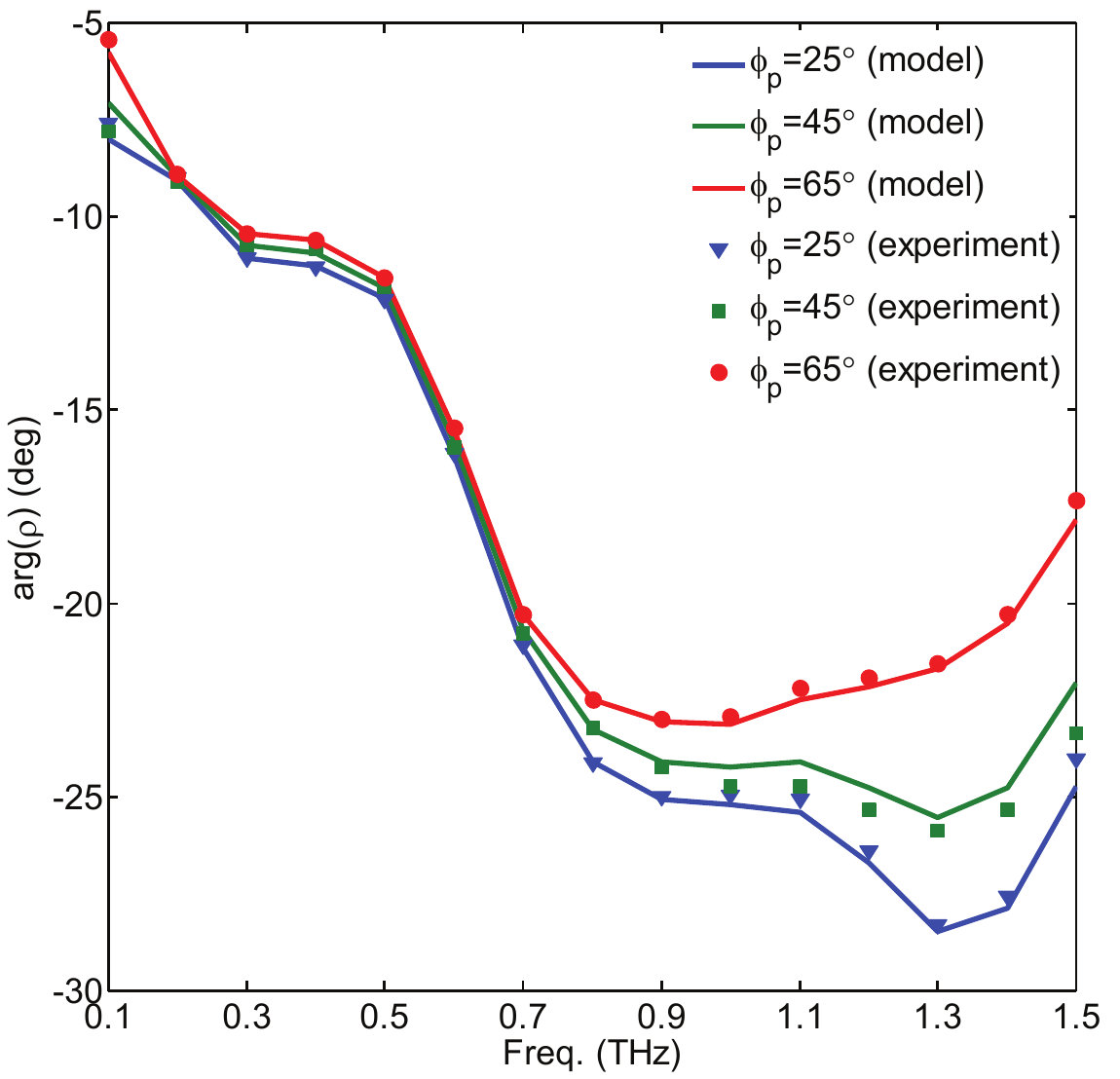}
}
\caption{Fitting results of the (a) amplitude and (b) phase of the $\rho$-parameter shown as solid lines for $\rho_{\textrm{mod}}$ and discrete marks for $\rho_{\textrm{exp}}$. The fitting was performed simultaneously for three sets of data corresponding to three different azimuthal angle of the polarizer and over the displayed frequency range. The sample was a highly doped silicon substrate, and the incidence angle was $73^\circ$.}\label{fitting}
\end{figure}

Figure \ref{fitting} shows the simultaneous fitting results of the amplitude and phase of the $\rho_{\textrm{mod}}$ to $\rho_{\textrm{exp}}$ for the highly doped silicon sample over the frequency range 0.1-1.5 THz. The fitting was performed simultaneously for three sets of data corresponding to three different azimuthal angle of the polarizer. Ellipsometric parameters $\Psi$ and $\Delta$ along with polarizer/analyzer extinction ratios, phase retardances and offset angles were the free parameters of the regression algorithm in this least squares fitting. This highly doped silicon sample is opaque in the THz range and cannot be measured in transmission. The incidence angle was $73^\circ$, which is close to the Brewster angle of silicon.

\begin{figure}[htbp]
\centering
\subfigure[]{
   \includegraphics[width=5.5cm] {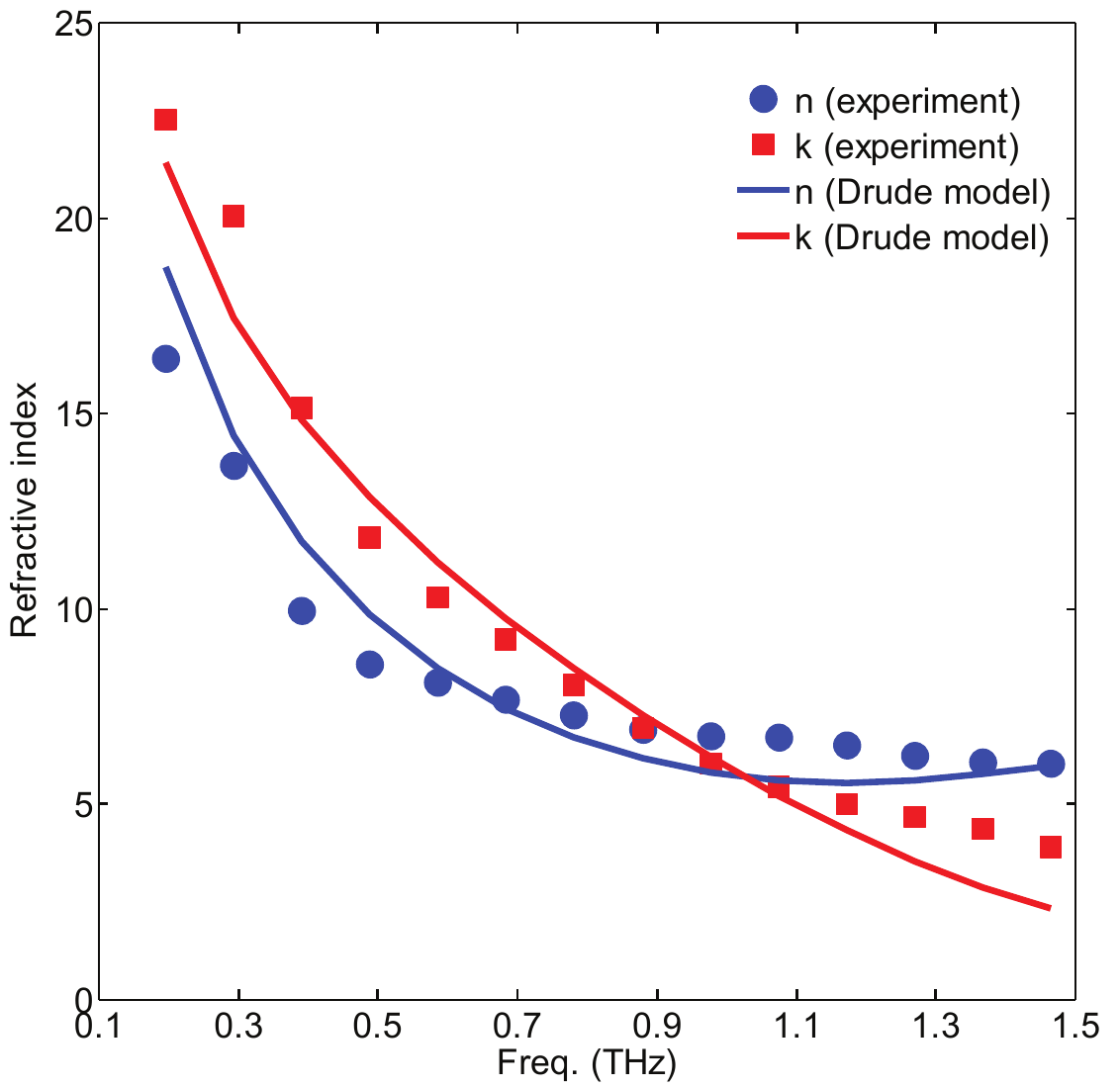}
}
\subfigure[]{
   \includegraphics[width=5.7cm] {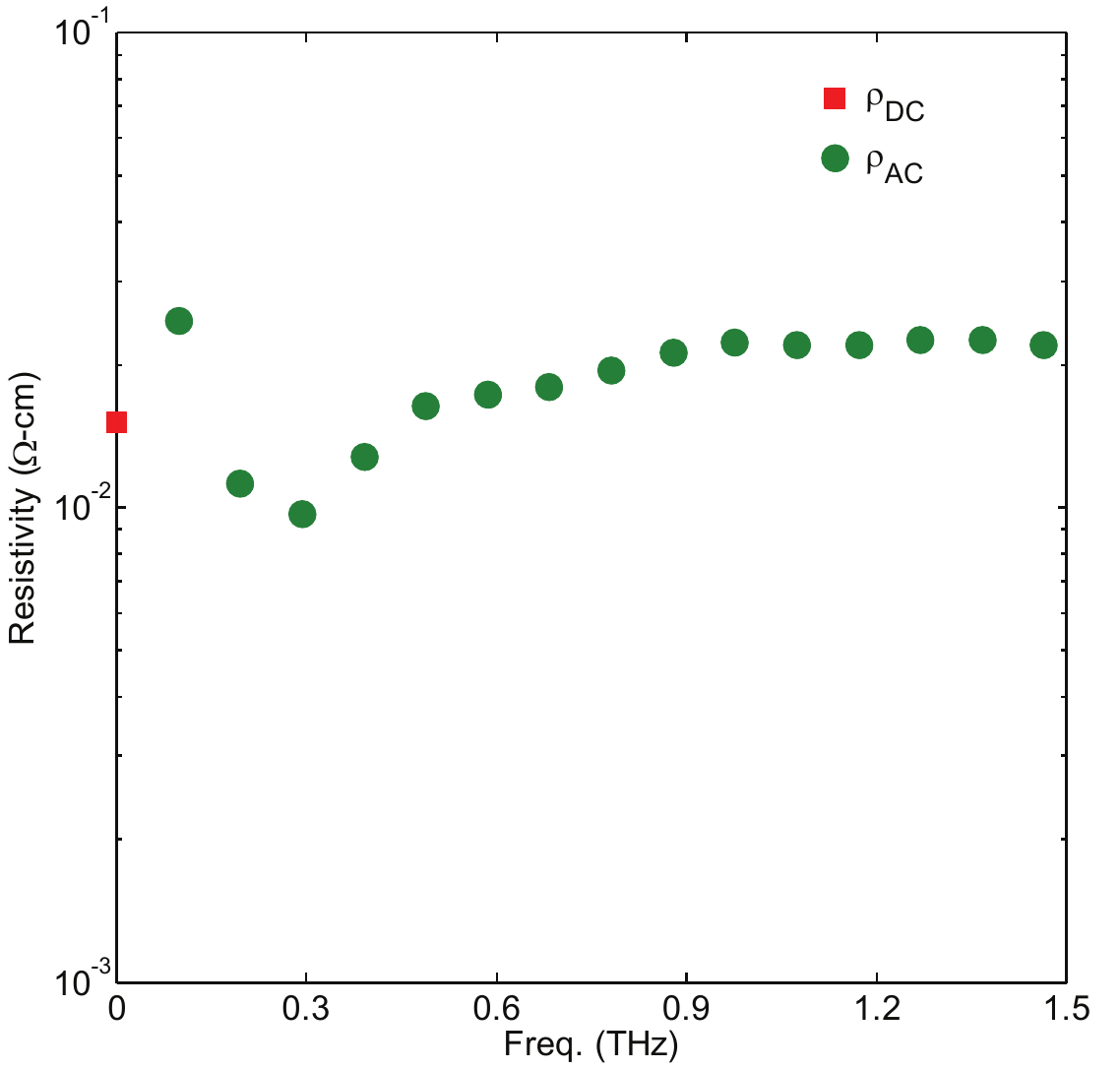}
}
\caption{(a) Extracted complex refractive index of the highly doped silicon substrate from ellipsometric parameters. Solid lines show the Drude model fit to the measured data.  (b) Extracted resistivity of the highly doped silicon substrate from ellipsometry measurement. Square mark presents its DC resistivity measured via a non-contact eddy-current resistivity gauge (COTS ADE 6035).}\label{Si0151}
\end{figure}

Figure \ref{Si0151}(a) shows the extracted complex refractive index of the highly doped silicon substrate from ellipsometric parameters along with a Drude model fit to the measured data. For the substrate resistivity and scattering time we obtain $\rho=0.011~\Omega$-cm and $\tau=190$ fs from the fitting and by assuming an effective electron mass of $0.26m_0$ \cite{Martin_1990}. These values are reasonable when compared with those previously reported in \cite{Hofmann_Feb2010} for a highly doped Si substrate. Figure \ref{Si0151}(b) shows the AC resistivity of the highly doped silicon substrate extracted from the ellipsometry measurement. As shown, the measured AC resistivity is consistent with the DC resistivity measured via a non-contact eddy-current resistivity gauge (COTS ADE 6035). However, there is dip in the resistivity curve around 0.3 THz. The exact reason of the dip is not clear to us at present, but we realized that polishing the substrate surface has considerable effect on its depth. Therefore, some surface effect may cause such feature in the resistivity curve.

\begin{figure}[htbp]
\centering
\subfigure[]{
   \includegraphics[width=5.7cm] {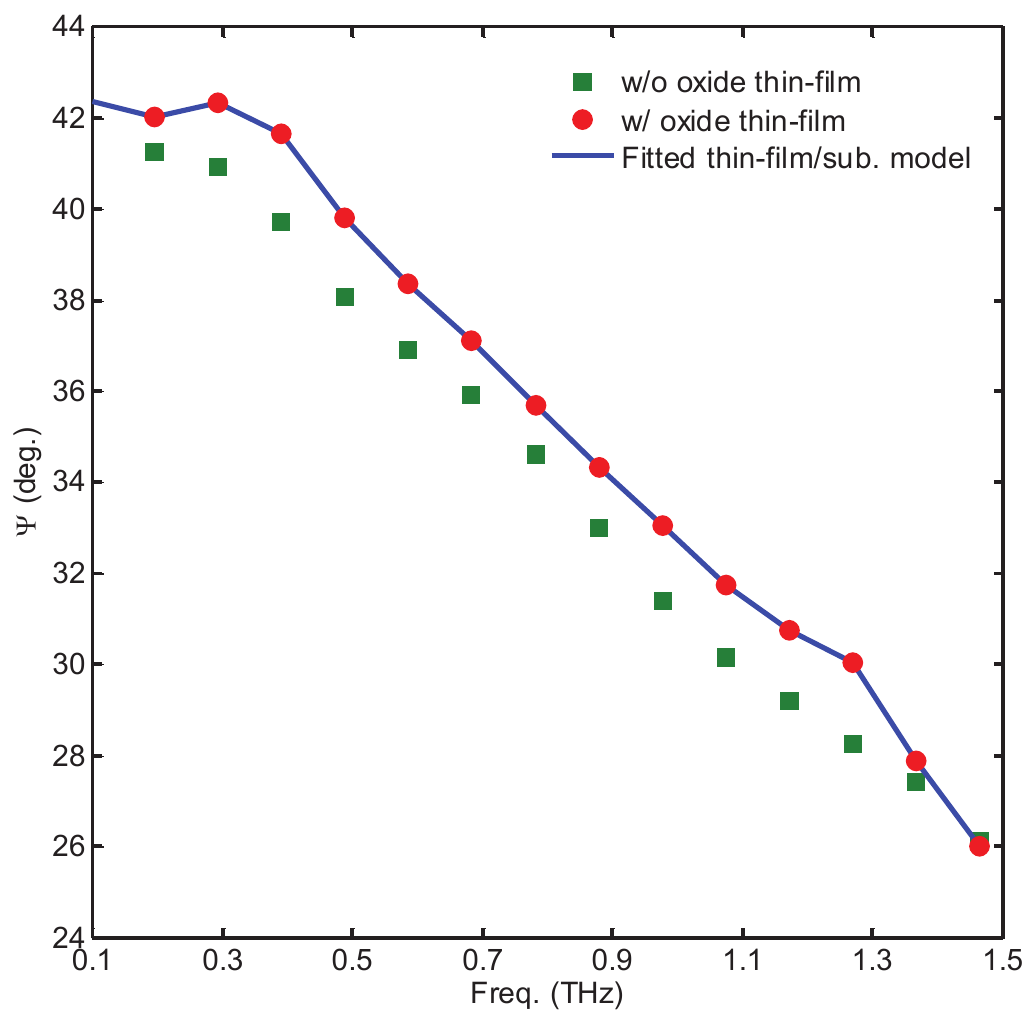}
}
\subfigure[]{
   \includegraphics[width=5.7cm] {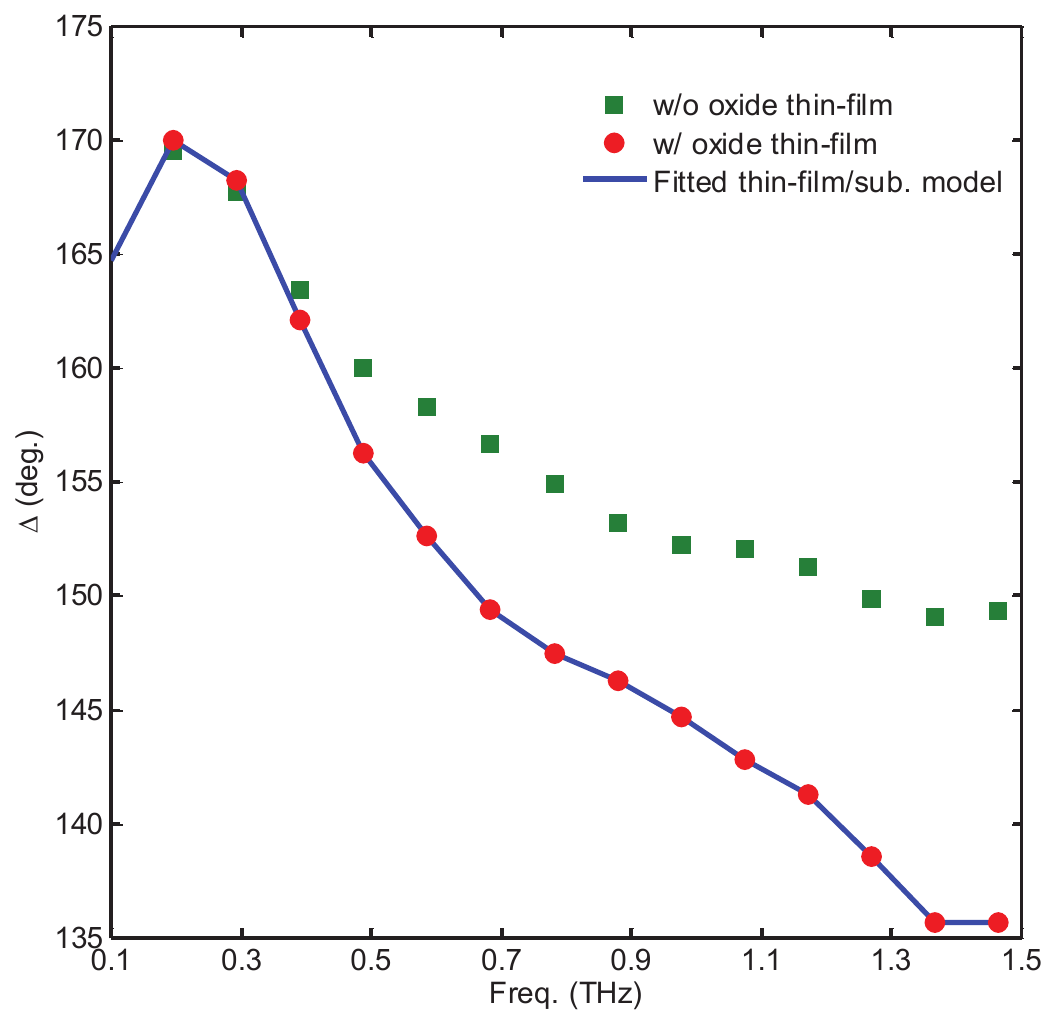}
}
\caption{Comparison between measured ellipsometric parameters (a) $\Psi$ and (b) $\Delta$ for the highly doped Si substrate w/ and w/o oxide thin-film. Solid line shows fitted thin-film/substrate model. The incidence angle was $73^\circ$.}\label{oxide}
\end{figure}

In order to further verify the overall accuracy of our THz-TDSE measurement method, we thermally grew a thin layer of oxide on top of the highly doped Si substrate. The thickness of the oxide thin-film was measured independently as 1.95 $\mu$m using optical techniques. Figure \ref{oxide} compares the measured ellipsometric parameters of the highly doped Si substrate before and after oxidization. The thickness of the oxide thin-film was extracted by fitting a \mbox{thin-film/substrate} model to the measured ellipsometric parameters. We used the same refractive index in Fig. \ref{Si0151}(a) for the substrate, and considered the refractive index of the thin-film as well as its thickness as free parameters in the fitting process. The resultant thickness of the oxide thin-film was obtained as 1.9 $\mu$m that is in excellent agreement with that obtained from the independent method. The fact that we can measure dielectric thicknesses that are only 0.3\% of the free space wavelength of 500 GHz radiation further shows the high accuracy of the technique. In order to avoid water vapor absorption, the space with terahertz wave propagation was enclosed and purged with dry air during the above measurements.

\section{Conclusion}
Terahertz time-domain spectroscopic ellipsometry is a newly established characterization technique that  collectively possesses the advantages of the well established techniques of ellipsometry and THz-TDS. In this work, a new experimental setup for THz-TDSE  was presented that easily provides variable angle of incidence in the range of $15^\circ$--$85^\circ$ in the reflection mode. The setup can also be configured into transmission mode with the same ease. We used hollow core photonic band gap fiber without pre-chirping to deliver femtosecond laser to the THz photoconductive antenna detector that facilitates the change of the incidence angle. An effective calibration scheme was proposed and applied for characterization of two different Si substrate and an oxide layer samples as an initial demonstration.

\section*{Acknowledgments}
This work was made possible by support from DARPA YFA N66001-10-1-4017 and the Gordon and Betty Moore Foundation. The authors would like to thank Mark A. Foster for his valuable input on the use of photonic band-gap fibers, Andrei Sirenko for discussions on far-infrared ellipsometry, and Jamie Neilson for helping with the SiO$_2$ growth.


\end{document}